\documentclass{aastex}
\newcommand{\hm}{H$_{2}$}

\newcommand{\nhp}{H$_{P}$}
\newcommand{\nhc}{H$_{C}$}

\newcommand{\php}{$\beta_{HP}$}
\newcommand{\phc}{$\beta_{HC}$}

\newcommand{\phhm}{$\beta_{H_{2}}$}
   
\begin{document}

\title{Molecular hydrogen formation in the interstellar medium}
\author{S.~Cazaux and A.G.G.M.~Tielens }
\affil{Kapteyn Astronomical Institute}
\affil{P.O.~Box 800, NL--9700 AV Groningen, The~Netherlands}
\email{Stephanie Cazaux (cazaux@astro.rug.nl)}

%%%%%%%%%%%%%%%%%%%%%%%%%%%%%%%%%%%%%%%%%%%%%%%%%%%%%%%%%%%%%%%%%
\begin{abstract}
We have developed a model for molecular hydrogen formation under
astrophysically relevant conditions. This model takes fully into
account the presence of both physisorbed and chemisorbed sites on the
surface, allows quantum mechanical diffusion as well as thermal
hopping for absorbed H-atoms, and has been benchmarked versus recent
laboratory experiments on \hm\ formation on silicate surfaces.  The
results show that \hm\ formation on grain surface is efficient in the
interstellar medium up to some 300K. At low temperatures ($\leq$100K),
\hm\ formation is governed by the reaction of a physisorbed H with a
chemisorbed H. At higher temperatures, \hm\ formation proceeds through
reaction between two chemisorbed H atoms. We present simple analytical
expressions for \hm\ formation which can be adopted to a wide variety
of surfaces once their surfaces characteristics have been determined
experimentally.
\end{abstract} 
 
\keywords{dust, extinction -- molecular hydrogen --ISM:molecules --
molecular processes }
%%     INTRODUCTION
%%     
\section{Introduction}
Molecular hydrogen is the most abundant molecule in the universe and
dominates the mass budget of gas in regions of star formation. \hm\ is
also an important chemical intermediate in the formation of larger
species and can be an important gas coolant in some conditions,
particularly in the early universe. Yet, despite its importance, the
\hm\ formation process is still not well understood. Observationally,
it has been shown that \hm\ can be efficiently formed over a wide
range of temperatures ( Jura 1974; Tielens \& Hollenbach 1985a, 1985b;
Hollenbach \& McKee 1979). Theoretically, Gould and Salpeter (1963)
showed the inefficiency of \hm\ formation in the gas phase and
concluded that recombination of physisorbed H on ``dirty'' ice
surfaces is efficient between 10 and 20K. Hollenbach and Salpeter
(1970; 1971), recognizing that this small temperature range presents a
problem, considered reactions involving H atoms bound to dislocations
and impurities with energy exceeding normal physisorbed energies, and
obtained a recombination efficiency $\simeq$ 1 up to a critical
temperature between 25 and 50K. Goodman (1978) calculated the quantum
and thermal mobility of the atoms on graphite grains assuming that
these atoms could only be physisorbed. Many studies focussed for
various reasons on icy surfaces where H is physisorbed (Buch \& Zhang
1991; Takahashi et al. 1999). However, most astrophysically relevant
material (e.g., silicates, graphite) can bind H also in chemisorbed
sites (Barlow \& Silk, 1976, Aronowitz \& Chang 1980; Leitch-Devlin \&
Williams 1984; Klose 1992; Fromherz et al. 1993). On these surfaces,
binding can occur in a physisorption layer (E$\sim$500K) at a distance
of some Z$\sim$3$\AA$ as well as in a chemisorption layer
(E$\sim$10000K) deeper into the surface Z$\sim$0.5$\AA$ (Barlow \&
Silk, 1976; Zangwill, 1988). Recently, Katz et al. (1999) developed a
model for \hm\ formation benchmarked by two sets of experiments. This
model considers the atoms bound only in physisorbed sites and
diffusing only thermally on the surface, colliding and recombining to
form molecules. As for the ice models studied earlier, this allows
molecule formation only below 15K for olivine grains and 20K for
carbon grains which contradicts the ISM observations. Perusing these
models, it is clear that the process of the \hm\ formation is governed
by the binding of atomic H to the surface and the concomitant mobility
of these atoms (Leitch-Devlin \& Williams 1984; Tielens \& Allamandola
1987). Surface diffusion can occur through quantum mechanical
tunneling (at low temperatures) and through thermal hopping (at high
temperatures). For a proper description of molecular hydrogen
formation in the ISM both types of binding diffusion processes have to
be taken into account (Cazaux \& Tielens 2002).

\section{Model}
Recently, we have developed a model for \hm\ formation on grain
surfaces based upon Langmuir kinetics where species accrete, migrate,
react and the product species, possibly assisted by thermal energy,
evaporate. This model is based on two main points: 1) The atomic
mobilities are due to a combination of quantum tunneling and thermal
diffusion and this is controlled by the temperature of the grain and
the mass of the species. 2) Atoms can bind to the surface in two
energetically different sites: a chemisorption site and a
physisorption site. These interaction energies set the migration
between the different sites and the reactions among the species. We
have compared the results of our model with laboratory experiments on
molecular hydrogen formation on olivine surfaces (Pirronello et
al. 1999; Katz et al. 1999) to determine the relevant
parameters. Here, we extrapolate this model to study \hm\ formation
under astrophysically relevant conditions (e.g., low accretion rates
and long timescales). We consider three different populations:
physisorbed H, chemisorbed H and physisorbed \hm. The surface
concentrations of these species are described by three rate
equations. Using the surface concentration of these species, the \hm\
desorption rate can be determined, which contains two
contributions. First order desorption occurs when the grain
temperature is high enough to allow evaporation of previously formed
molecules. Second order desorption occurs when two atoms encounter and
the newly formed molecule is directly released into the gas phase. We
define $H_{P}$, $H_{C}$ and $H_{2}$ as the physisorbed H, chemisorbed
H and molecular hydrogen concentrations, respectively; their
evaporation rate are written as \php, \phc, \phhm. The different
mobilities to go from a site $i$ to a site $j$ are given by
$\alpha_{ij}$ where $i$ and $j$ are either a physisorbed site (P) or a
chemisorbed site (C), and where the barrier between two sites is
assumed to be square. However, the shape of the barrier is found out
to be unimportant (see below). $\mu$ is the fraction of \hm\ which
stays on the surface upon formation. The rate equation can then be
written as (for details see Cazaux \& Tielens 2002),

\begin{equation}
\dot{H}_{P}=F(1-H_{P}-{H_{2}})-\alpha_{pc}H_{P}-2\alpha_{pp}{H_{P}}^2+\alpha_{cp}H_{C}(1-H_{P})-\beta_{H_P}H_{P}
\end{equation}
\begin{equation}
\dot{H}_{C}=\alpha_{pc}H_{P}(1-H_{C})-\alpha_{pc}H_{P}H_{C}-\alpha_{cp}H_{C}-2\alpha_{cc}{H_{C}}^2\beta_{H_C}H_{C}
\end{equation}
\begin{equation}
\dot{H}_{2}=\mu(2\alpha_{pp}{H_{P}}^{2}+\alpha_{pc}H_{P}H_{C}+\alpha_{cp}H_{C}H_{P}+\alpha_{cc}{H_{C}}^2)-\beta_{H_2}H_{2}
\end{equation}

Where F is accretion rate in unit of monolayers per second
(ML.$s^{-1}$). The first order \hm\ desorption rate is given by:
\begin{equation}
k_1=\beta_{H_{2}}H_{2}
\end{equation}

For second order desorption several processes contribute depending on
whether physisorbed, chemisorbed atoms or both are involved:

\begin{equation}
k_2=(1-\mu)(\alpha_{pc}H_{P}H_{C}+\alpha_{cp}H_{C}H_{P}+\alpha_{p}H_{P}^2+\alpha_{cc}H_{C}^2)
\end{equation}

$\beta_i$, the desorption rate for a population $i$ is written
$\beta_i=\nu_{i} \exp\left[-\frac{E_{i}}{k T}\right]$, where $\nu_{i}$
is the frequency factor of species $i$, perpendicular to the surface,
and $E_i$ the desorption energy of species $i$. The total \hm\
formation rate, R, is the sum of the first and second order terms, and
the \hm\ recombination efficiency -- the fraction of the accreted
hydrogen which desorbs as \hm -- is defined by $\epsilon=\frac{2R}{F}$
. The parameters in this model are summarized in Table
~\ref{table}. They have been discussed extensively elsewhere (see
Cazaux \& Tielens 2002). We have integrated this set of time dependent
equations using a Runge-Kutta method with adaptative stepsize control,
(Odeint, Subroutine package from Numerical Recipes 1992) until steady
state was achieved. Under astrophysically relevant conditions, steady
state is generally achieved. These steady state results are shown as a
function of T in fig.~\ref{table} for three different H-fluxes. All
curves are characterized by a very low efficiency at low temperatures,
a steep rise around $\sim$5K -- whose location depends on the actual
H-flux --, an efficiency of unity until some 25K, followed by a
gradual decline to $\sim$ 0.2 at $\sim$ 100K, and a final drop to very
low efficiency around 300K, whose location again depends on the actual
H accretion rate. In steady state, the system of equations 1-3 can be
simplified by setting the time derivatives equal to zero. We can now
discern two different temperatures regimes -- low temperatures
(T$\leq$100K) and high temperatures (T$\geq$100K) -- where different
processes dominate \hm\ formation. In these limits, the set of steady
state equations can be further simplified and this yields much insight
in the numerical results.

\subsection{Physisorbed H}
 At low temperatures, by far the most efficient way to form \hm\ is
 when a physisorbed atom reacts with a chemisorbed atom. Other routes
 forward \hm\ as well as the evaporation of chemisorbed atoms are
 negligible. Realizing that $H_P$ is always small compare to 1
 (physisorbed H quickly migrate to a chemisorbed well), the steady
 state equations can be simplified to,
\begin{equation}
F(1-H_2)-\alpha_{pc}H_P-\beta_{H_{P}}{H_{P}}=0
\end{equation}
\begin{equation}
\alpha_{pc}H_P(1-2H_{C})=0
\end{equation}
\begin{equation}
\mu(\alpha_{pc}H_P H_C)-\beta_{H_2}H_2=0
\end{equation}

The different populations can easily be determined in this system
(e.g., $H_C$=$\frac{1}{2}$,
$H_P$=$(\frac{\mu\alpha_{pc}}{2\beta_{H_2}}+\frac{\alpha_{pc}}{F}+\frac{\beta_{H_2}}{F})^{-1}$
and $H_2$=$\frac{\mu\alpha_{pc}}{2\beta_{H_2}}H_P$), the \hm\
desorption rate is calculated and we deduce the recombination
efficiency:
\begin{equation}
\epsilon_{H_2}=\left({\frac{\mu F}{2\beta_{H_{2}}}}+1+{\frac{\beta_{H_{P}}}{\alpha_{pc}}}\right)^{-1}
\end{equation}
We can recognize three different regimes in this temperature range. At
really low temperatures, below some 10K, the high mobility of H atoms
due to tunneling permits the recombination of incoming H
atoms. The \hm\ formed stays on the surface, since the temperature is
not high enough to allow evaporation and blocks further H-atom
accretion. Then, the recombination efficiency is
$\epsilon_{H_2}$=$\left[{\frac{\mu
F}{2\beta_{H_{2}}}}\right]^{-1}$. At higher temperatures, (6-25K) the
desorption rate depends only on the flux because all the incoming H
atoms adsorb, recombine and desorb as \hm. In this temperature regime,
hydrogen accretes into a physisorbed site but quickly drops into a
chemisorbed well. If this chemisorbed site is empty, the H atom will
be trapped. If an H atom is already present, reaction will occur and
the product will evaporate quickly either upon formation or thermally
assisted; thus, $H_C$=$\frac{1}{2}$ and $\epsilon_{H_2}$=1. Between
$\sim$25K up to $\sim$100K, evaporation of physisorbed atoms competes
with recombination, and the desorption rate decreases
considerably.
\subsection{Chemisorbed H}
With increasing temperature, the physisorbed (at T$\sim$100K) and then
the chemisorbed atoms(at T$\sim$300K) will start to evaporate. The
most efficient reaction to form \hm\ is then the ``collision'' of two
chemisorbed atoms. The system of equations reduces to,
\begin{equation}
F-\alpha_{pc}H_P-\beta_{H_{P}}{H_{P}}=0
\end{equation}
\begin{equation}
\alpha_{pc}H_P-2\alpha_{cc}H_C^2-\beta_{H_C}{H_C}=0
\end{equation}
\begin{equation}
\mu(\alpha_{cc}{H_C}^2)-\beta_{H_{2}}H_2=0
\end{equation}

and the \hm\ recombination efficiency can be written as,

\begin{equation}
\epsilon=\left({1+{\frac{\beta_{H_P}}{\alpha_{pc}}}}\right)^{-1}\xi
\end{equation}
with $\xi$ the correction factor at high temperatures, 
\begin{equation}
\xi= \left({1+{\frac{\beta_{H_C}^2\beta_{H_P}}{2F\alpha_{pc}\alpha_{cc}}}}\right)^{-1},
\end{equation}
which reflects the evaporation of chemisorbed H. This factor is flux
dependent.

\subsection{General expression for the recombination efficiency}
The above expression for the \hm\ recombination efficiency can be
combined into one general expression valid at any temperatures:

\begin{equation}
\epsilon_{H_2}= \left({{1+{\frac{\mu F}{2\beta_{H_2}}}+{\frac{\beta_{H_P}}{\alpha_{pc}}}}}\right)^{-1}\xi\\
\end{equation}
The mobilities, $\alpha_{PC}$ and $\alpha_{CC}$, are due to a
combination of quantum mechanical tunneling and thermal hopping but
considering the temperatures at which these parameters play a role,
these mobilities are dominated by thermal hopping. Therefore, the \hm\
recombination efficiency is independent of the width of the considered
barriers, and we can approximate $\xi$ and
${\frac{\beta_{H_P}}{\alpha_{pc}}}$ by the following expressions:

\begin{equation}
\xi=\left({1+\frac{\nu_{H_C}\exp\left[-\frac{1.5E_{H_C}}{kT}\right]\left(1+\sqrt{\frac{E_{H_C}-E_S}{E_{H_P}-E_S}}\right)^2}{2F}}\right)^{-1}
\end{equation}

\begin{equation}
{\frac{\beta_{H_P}}{\alpha_{pc}}}=\frac{1}{4}\left(1+\sqrt{\frac{E_{H_C}-E_S}{E_{H_P}-E_S}}\
\right)^2 \exp\left[-\frac{E_S}{kT}\right]
\end{equation}

\begin{figure}
\plotone{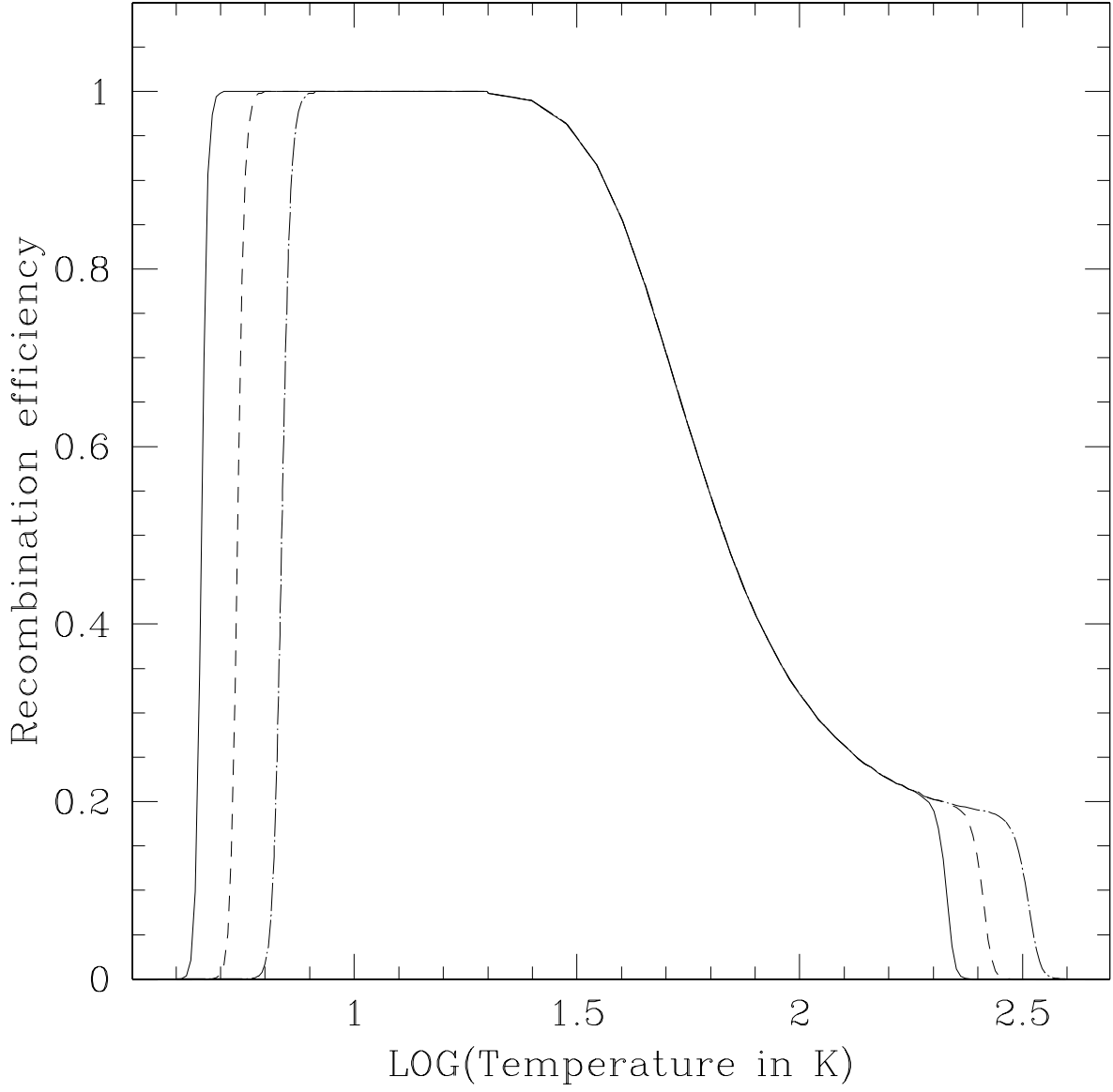}
\caption{\hm\ recombination efficiency for three different flux:
$10^{-15}$ monolayers per second (solid), $10^{-10}$ monolayers per
second(dash) and $10^{-5}$ monolayers per second (dot-dash). }
\label{pap}
\end{figure}

The parameters in these expressions have been determined from
experimental data (Table ~\ref{table}; Cazaux \& Tielens 2002). The
\hm\ recombination efficiency for three different flux is reported
figure ~\ref{pap}. In astrophysical environments, the recombination
rate is written:
\begin{equation}
R_{H_2}={\frac{1}{2}} n_H v_H n_d \sigma_d \epsilon_{H_2} S_{H}(T)
\end{equation}
where $n_H$ and $v_H$ are the number density and the thermal velocity
of H atoms in the gas phase, $n_d \sigma_d$ is the total cross section
of the interstellar grains and $S_{H}(T)$ is the sticking
coefficient of the H atoms which can depend on temperature.

\section{Discussion}
Our study reveals the presence of two distinct regimes of \hm\
formation, which reflect directly the presence of two types of atomic
H binding sites. At low temperature (T$\leq$100K), \hm\ formation
involves migration of physisorbed H atoms. At higher temperatures
(T$\geq$100K), \hm\ formation results from chemisorbed H
recombination.  The presence of these two types of binding sites allow
\hm\ formation to proceed relatively efficiently even at elevated
temperatures. The study of Hollenbach and Salpeter (1971) focused on
icy surfaces on which H can only physisorb. As a result, \hm\
formation ceased at temperatures in excess of $\sim$20K. Recognizing
this problem, Hollenbach and Salpeter involved the presence of
enhanced binding sites on the ice with ill-determined
parameters. These sites allowed \hm\ formation to proceed up to some
75K. Since their study, it has become abundantly clear that
interstellar grains are not covered by ice in the diffuse interstellar
medium (Whittet et al. 1983, 1988). Silicate and graphitic surfaces
are now widely accepted as astrophysically relevant grain surfaces
(Mathis 1991) and those surfaces intrinsically possess enhanced
binding sites; e.g., chemisorbed sites. The parameters of these
chemisorbed sites have not yet been well determined because
experiments have focused on low temperature \hm\ formation (Pirronello
et al. 1997; Katz et al. 1999). The values adopted in this study are
however quite reasonable and illustrate the efficiency of \hm\
formation at elevated temperatures well. When future experiments
determine the values of the parameters involved ($E_{H_C}$, $E_{H_P}$,
$\mu$), the results can be directly adjusted. Inside dense clouds,
interstellar grains are covered by ice. Of course, in such
environments, almost all hydrogen is already in molecular form and
\hm\ formation is perhaps only of academic interest. Nevertheless, we
note that in such environments, molecular hydrogen formation may
proceed mainly through H-abstraction from molecules such as H$_{2}$S
and N$_{2}$H$_{2}$ (Tielens and Hagen 1982).  In a sense, these
species act as ``chemisorption'' sites for hydrogen. Migrating atomic
H may tunnel through the reaction barriers involved and form
\hm. Eventually, these ice covered grains are transported back into
the diffuse ISM, when the molecular cloud is
disrupted. Photodesorption and sputtering in strong shocks quickly
remove their ice on a timescale of some $10^6$ yr (Tielens and Hagen
1982, Jones et al. 1994, Draine and Salpeter 1979, Barlow 1978). At
that point, molecular hydrogen formation is again governed by the
properties of bare grain surfaces. Similarly, any (thin) layer
accreted in the diffuse interstellar medium will be quickly sputtered
in even a modest velocity ($\sim$30km s$^{-1}$) shock (Jones et
al. 1994).\\Finally, the formation efficiency of molecular hydrogen
will also depend on the sticking coefficient of H atoms colliding with
the grain. In our model and the formulae derived, the sticking
coefficient is subsumed in the incident flux, F. Astrophysical studies
of sticking of H on grain surfaces have concentrated on physisorbed
interactions and the sticking coefficient is $\sim$1 at low
temperatures and decreases with increasing temperature to about 0.3 at
T=300K ( Hollenbach \& Salpeter 1970; Hollenbach \& McKee 1979; Burke
\& Hollenbach 1983; Leitch-Devlin \& Williams 1985). However, if the
interaction occurs through much stronger chemisorption, then the
sticking coefficient might be large even at high temperatures (Tielens
\& Allamandola 1987; Duley \& Williams 1984).

\subsection{Summary and conclusions}
Recently, we have modelled molecular hydrogen formation on grain
surfaces. This model consider hydrogen atoms bound to the surface at
two energy levels (i.e. chemisorption and physisorption). The H
mobility from one site to another is a combination of tunelling effect
and thermal diffusion. This model has been experimentally benchmarked
(Pirronello et al. 1997a, 1997b, 1999) and the relevant surface
characteristics have been determined. These characteristics allow us
to extend our model for \hm\ formation under astrophysically relevant
conditions. The results show efficient \hm\ formation from $\sim$6K to
$\sim$300K. The different processes involved in \hm\ formation at
different temperatures has been discussed.  Until about 100K, \hm\
forms by recombination of a physisorbed H with a chemisorbed H and is
highly efficient. At higher temperatures, when physisorbed atoms
evaporate quickly, the recombination of two chemisorbed atoms is
required to form \hm. \hm\ formation is then less efficient,
$\epsilon_{H_2}$$\sim$0.2. The parameters involved in H chemisorption
and \hm\ formation at high temperatures are presently not well
known. The adopted values are very reasonable and the gross
characteristics -- \hm\ formation at high temperatures -- are
undoubtedly correct. Nevertheless, future experiments are very
important to determine the maximum temperature to which \hm\ formation
in the ISM can occur.

\clearpage

\begin{table}
\caption{Model parameters for silicate surface.\label{table}} 
%\begin{center}
\begin{tabular}{crrrrrr}
\tableline\tableline \multicolumn{1}{c}{$E_{H_2}$\tablenotemark{a}} &
\multicolumn{1}{c}{$\mu$\tablenotemark{a}} &
\multicolumn{1}{c}{$E_{S}$\tablenotemark{a}} &
\multicolumn{1}{c}{$E_{H_P}$\tablenotemark{a}} &
\multicolumn{1}{c}{$E_{H_C}$\tablenotemark{a}} &
\multicolumn{1}{c}{$\nu_{H_2}$\tablenotemark{a}} &
\multicolumn{1}{c}{$\nu_{H_C}$\tablenotemark{a}} \\ \tableline
\multicolumn{1}{c}{K} & & \multicolumn{1}{c}{K} &\multicolumn{1}{c}{K}
&\multicolumn{1}{c}{K}& \multicolumn{1}{c}{$s^{-1}$}&
\multicolumn{1}{c}{$s^{-1}$} \\ 320&0.005&200&600&$~$10000&3 10$^{12}$&1.3 10$^{13}$\\
\tableline
\end{tabular}
\tablenotetext{a}{for more details about the determination and
calculation of these parameters, see Cazaux \& Tielens 2002}
\tablecomments{ $E_{H_2}$, $E_{H_P}$ and $E_{H_C}$ are the desorption
energies of \hm, physisorbed H (\nhp) and chemisorbed H (\nhc), and
$E_S$ is the energy of the saddle point between two physisorbed
sites. $\mu$ is the fraction of the newly formed \hm\ which stays on
the surface and $\nu_{H_2}$ and $\nu_{H_C}$ are the vibrational
frequencies of \hm\ and H in their surface sites.}
%\end{center}
\end{table}

\end{document}